\documentstyle[aps,prl,amsfonts,amssymb,amsmath,graphicx]{revtex}

\begin{document}
\def\BJ{{\mathcal{J}}}
\def\u{\boldsymbol}
\draft
\title{Stochastic ionization through noble tori~: Renormalization results }

\author{C. Chandre and T. Uzer}
\address{Center for Nonlinear Science, School of Physics, Georgia
Institute of Technology, Atlanta, Georgia 30332-0430}
\date{\today} \maketitle

\begin{abstract}
We find that chaos in the stochastic ionization problem develops through the
break-up of a sequence of noble tori.  In addition to being very accurate, our
method of choice, the renormalization map, is ideally suited
for analyzing properties at criticality. Our computations of chaos thresholds
agree closely with the widely used empirical Chirikov criterion. \end{abstract}
\pacs{PACS: 32.80.Rm, 05.45.Ac, 05.10.Cc}

The multiphoton ionization of hydrogen in a strong microwave field~\cite{bayf74}
revolutionized traditional notions about the physics of highly excited
atoms. Its interpretation remained a puzzle until its stochastic,
diffusional nature was uncovered through the then-new theory of
chaos~\cite{meer79} thereby making it the most important
testing ground for quantal manifestations of classical chaos~\cite{casa87}.
The onset of chaos in that problem has been the focus of intensive research and
is generally believed to be induced by the break-up of invariant tori
~\cite{Breic92}, which act as barriers in phase space preventing the diffusion of
trajectories for Hamiltonian systems with two degrees of freedom.

Over the last two decades, different methods to estimate break-up
thresholds of invariant tori have been developed and applied to
various physical models with two effective degrees of freedom: Chirikov's
resonance overlap criterion~\cite{chir79}, its empirical ``2/3-rule''
version~\cite{Blich83}, and Escande-Doveil renormalization~\cite{esca85}. These
methods provide merely approximate values (as compared to direct numerical
integration) even though Escande-Doveil
renormalization is known to give very accurate values in some concrete
examples. Systematic methods are now available to obtain very accurate
values, such as, e.g., Greene's residue criterion~\cite{gree79},
Laskar's frequency map analysis~\cite{lask93,lask99}, or
renormalization analysis~\cite{chan01PR,chan98b,chan00a,chan01a}.
There is numerical evidence that these independent and systematic
methods give the same values for the chaos
thresholds~\cite{chan98b,chan01a,chan00d}. However, renormalization
has an additional advantage~:
By focusing on specific tori, it leads to very
accurate thresholds, and allows one to analyze the properties at criticality.
Indeed, the renormalization map can be likened to a phase space microscope
by which the system can be studied with larger and larger magnification.

In this article, we find the transition to chaos in the hydrogen atom driven by
circularly polarized microwaves using renormalization. This problem has emerged
as paradigm for a number of issues in multi-dimensional nonlinear dynamics (see
Ref.~\cite{brun97} and references therein). Our results reveal that the onset of
chaos in the microwave problem develops through a sequence of ``noble'' tori,
i.e.\ tori with frequency equivalent to the golden mean $\gamma=(\sqrt{5}-1)/2$.

Furthermore, the chaos thresholds obtained by the renormalization are in very good
agreement with the 2/3-rule
criterion~\cite{delo83,howa92,sach97}, and
better than the ones obtained by the Escande-Doveil
renormalization~\cite{sach97}. In this model, Chirikov's criterion
can be used to determine very accurate critical thresholds of
ionization (accurate to 5\%) for most values of the parameter (eccentricity of
the initial orbit).

\paragraph*{Renormalization method--}

The renormalization method is based on the construction of successive
canonical transformations~\cite{chan01PR,chan00a,chan01a}. In its most recent 
version, it acts on the following family of Hamiltonians with
two degrees of freedom written in actions $\u{A}=(A_1,A_2)$ and angles
$\u{\varphi}=(\varphi_1,\varphi_2)$~:
\begin{eqnarray}
H(\u{A},\u{\varphi}) &=&  \u{\omega}\cdot\u{A}+V(\u{\Omega}\cdot \u{A},\u{\varphi})
\label{eqn:form}\\
&=&\u{\omega}\cdot\u{A}+\sum_{\u{\nu}\in{\mathbb{Z}}^2\atop k\in
{\mathbb{N}}} V_{k,\u{\nu}}(\u{\Omega}\cdot\u{A})^k
e^{i\u{\nu}\cdot\u{\varphi}}, \nonumber \end{eqnarray} where
$\u{\omega}=(\omega,-1)$ is the frequency of the invariant torus, and
$\u{\Omega}=(1,\alpha)$ is some other vector.
Two steps are involved~\cite{koch99}~:
an elimination of the non-resonant modes $\u{\nu}$ of
the Hamiltonian (the modes which do not involve small denominator problems), and
a rescaling of phase space (shift of the resonances, rescaling in time and in
the actions). This transformation reduces to a map ${\mathcal{R}}$
of Fourier coefficients
$(\omega',\alpha';\{V'_{k,\u{\nu}}\})={\mathcal{R}}(\omega,\alpha;\{V_
{ k , \u{\nu}}\})$.

 The main conjecture
of the renormalization approach is that if the
torus exists for a given Hamiltonian $H$, the iterates ${\mathcal{R}}^n H$
of the renormalization map acting
on $H$
converge to some integrable Hamiltonian $H_0$. This
conjecture is supported by analytical results in the perturbative
regime~\cite{koch99,koch99b}, and by numerical results~\cite{chan00d,chan01a}.
For a one-parameter family of Hamiltonians
$\{H_F\}$, the critical amplitude of the perturbation
$F_c(\omega)$ is determined by the following conditions~:
\begin{eqnarray} && {\mathcal R}^nH_{F} \underset{n\to \infty}{\to}
H_0(\u{A}) =\u{\omega}\cdot\u{A}+ \frac{1}{2}(\u{\Omega}\cdot\u{A})^2
\quad \mbox{ for } F<F_c(\omega), \label{eqn:Adef1}\\ && {\mathcal
R}^nH_{F} \underset{ n\to \infty}{\to} \infty \quad \mbox{ for }
F>F_c(\omega).\label{eqn:Adef2} \end{eqnarray}

\paragraph*{Model--}

We consider a hydrogen atom interacting with a strong microwave field of
amplitude $F$ and frequency $\Omega$, circularly polarized
in the orbital plane.
In action-angle variables, the classical Hamiltonian is reduced to the following
Hamiltonian with
two degrees of freedom~\cite{howa92}~: \begin{equation}
\label{eqn:ham}
H(J,L,\theta,\psi)=-\frac{1}{2J^2}-\Omega L+F\sum_{n=-\infty}^{+\infty}
V_n(J,L)\cos (n\theta+\psi),
\end{equation}
where
\begin{eqnarray*}
&& V_0(J,L)=-\frac{3e}{2}J^2,\\
&& V_n(J,L)=\frac{1}{n}\left[ \BJ'_n(ne)+
\frac{\sqrt{1-e^2}}{e}\BJ_n(ne)\right]
J^2, \, \mbox{ for } n\not= 0,
\end{eqnarray*}
where $\BJ_n$ is the $n$th Bessel function of the
first kind and $\BJ_n'$ is its derivative. The angles $\theta$ and $\psi$
are conjugate to the principal action $J$ and to the
angular momentum $L$ respectively. The eccentricity $e$ of the initial orbit is
given by $e=(1-L^2/J^2)^{1/2}$.
Hamiltonian~(\ref{eqn:ham}) can be rescaled in order to eliminate
the dependence on the frequency of the microwave field. We rescale
time by a factor $\Omega$ [we divide Hamiltonian~(\ref{eqn:ham}) by
$\Omega$]. We rescale the actions $J$ and $L$ by a factor
$\lambda=\Omega^{1/3}$, i.e.\ we replace $H(J,L,\theta,\psi)$ by
$\lambda H(J/\lambda,L/\lambda,\theta,\psi)$. We notice that this
rescaling does not modify $e(J,L)$. The resulting
Hamiltonian becomes~: $$
H=-\frac{1}{2J^2}-L+F'\sum_{n=-\infty}^{+\infty} V_n(J,L)\cos
(n\theta+\psi), $$
where $F'=F\Omega^{-4/3}$ is the rescaled amplitude of the field. In
what follows we assume that $\Omega=1$.

For the unperturbed Hamiltonian~(\ref{eqn:ham}), i.e.\ with $F=0$, we
consider a motion with Kepler frequency $\omega\in [-1,1]$ (the high scaled
frequency regime).
In phase space, this trajectory evolves on a
two-dimensional torus. If $\omega$ is irrational, the trajectory fills
the torus densely. This invariant torus is located at
$J=\omega^{-1/3}$ and $L=-E_0-\frac{1}{2}\omega^{2/3}$, where $E_0$ is
the total energy of the system (in the rotating frame). For
convenience, we shift the action $J$ such that the torus is located at
$J=0$, and we expand the resulting Hamiltonian in Taylor series in the
action $J$. Furthermore, we rescale the actions $J$ and $L$ by a
factor $\lambda=-3\omega^{4/3}$, making Hamiltonian~(\ref{eqn:ham})~:
\begin{equation}
\label{eqn:hamresc}
H_F=\omega J -L
+\sum_{k=2}^{+\infty} \frac{k+1}{6(3\omega)^{k-2}}J^k +F\left(
-\frac{1}{3}\omega^{-4/3} J^2 +2\omega^{-1/3} J -3\omega^{2/3}\right)
\sum_{n=-\infty}^{+\infty} \tilde{V}_n(e)\cos(n\theta+\psi),
\end{equation}
where $\tilde{V}_n(e)=V_n(1,\sqrt{1-e^2})$. We note
that the rescaling coefficient $\lambda$ has been chosen such that
the quadratic part of the Hamiltonian is $J^2/2$.
The resulting Hamiltonian conforms to~(\ref{eqn:form}) with $\alpha=0$
and with
coordinates $\u{A}=(J,L)$ and $\u{\varphi}=(\theta,\psi)$.

In what follows, we consider $e$ (the eccentricity of the initial orbit)
as a parameter of the system. For $e=1$,
the resulting model is a one-dimensional hydrogen atom in a {\em
linearly} polarized microwave field~\cite{casa87}. By varying $e$,
we obtain a wide variety of models where we can compare renormalization and
empirical rules.

\paragraph*{Breakup thresholds of invariant tori--}

For a given eccentricity $e$, we compute the critical function
$F_c(\omega ; e)$ using the renormalization method
[Eqs.~(\ref{eqn:Adef1})-(\ref{eqn:Adef2})].
Figure~1 shows a typical critical function $\omega\mapsto
F_c(\omega;e)$ for $e=0.45$ and for co-rotating orbits ($\omega >0$).
The critical function vanishes at all rational values of the frequency
(since all tori with rational frequency are broken as soon as the
field is turned on), and it is discontinuous on a dense set. This figure shows
that there is no invariant torus between the resonances 1:1
and 4:1 for $F_c>\max_{\omega\in[1/4,1]}F_c(\omega;e=0.45)\approx
0.0069$. The most stable region is located around $\omega \approx
0.723$.

Figure 2 shows the critical thresholds between two primary resonances
$\max_{\omega\in [1/2,1]} F_c(\omega,e)$ and $\max_{\omega\in [1/3,1/2]}
F_c(\omega,e)$ as a function of the parameter $e$ for co-rotating ($\omega >
0$) and for counter-rotating ($\omega<0$) orbits. This figure is
analogous to Fig.~3 of Ref.~\cite{sach97}. In a broad range of values of
$e$ ($e\in [0.2,1]$) the heuristic 2/3-rule criterion gives very
accurate results (accurate to 5\%) by comparison with our renormalization
results for co-rotating orbits. The values are even better than the
ones computed by Escande-Doveil renormalization~\cite{sach97}.
However, for very low eccentricity,
there is a large gap between both results which makes the 2/3-rule
inapplicable in this range of parameter $e$. For this case, Escande-Doveil
renormalization is a better criterion for the determination of the thresholds.
As $e$ tends to zero, the discrepancy is even bigger. For instance, for the
overlap between $1:1$ and $2:1$, the 2/3-rule criterion predicts a finite value of
the threshold at $e=0$ whereas Hamiltonian~(\ref{eqn:hamresc}) is integrable in that case
and the critical function is expected to go to infinity at $e=0$.

For counter-rotating
orbits, the 2/3-rule criterion overestimates the critical couplings
even though it gives fairly accurate results (less than 10\% for $e\in
[0.8,1]$). The resulting critical curve is below what has been
obtained in Ref.~\cite{sach97} using the Escande-Doveil
renormalization with a discrepancy around 30\%.

Our results confirm that the orbits with
medium eccentricity can diffuse more easily between the first primary
resonances ($n=1,2,3$)~\cite{sach97}. However, in order to ionize, the orbits must
diffuse throughout phase space or at least between a large number of
primary resonances $n_c^q$ (in experiments and numerical
simulations~\cite{bell96}, $n_c^q\approx 40$).
Since the 2/3-rule gives very accurate results in a
broad range of parameter $e$, we compute the critical thresholds
between resonances $m$:1 and $m$+1:1 for $m=1,\ldots,n_c^q$ with
$n_c^q=40$, for medium and large eccentricities. In Fig.~3 we have
plotted the different critical curves $e\mapsto \max_{\omega\in
[1/(m+1),1/m]} F_c(\omega;e)$ for $m=1,\ldots,n_c^q$ for co-rotating
and counter-rotating orbits. It appears that there is a very broad
region of the parameter $e$ where an orbit cannot diffuse from the
resonance 1:1 to $n_c^q$:1. In fact, only in the region $e\in[0.9,1]$
(with $F> 0.015$ for co-rotating orbits, or with $F>0.05$ for
counter-rotating ones), the orbits can ionize. In other terms, only
the high-eccentricity orbits can ionize in this classical
model~(\ref{eqn:hamresc}). This observation reinforces the importance
of core collisions for the ionization process~\cite{brun97}.

A more commonly used function is the scaled function
$F_0(\omega)=F_c(\omega)\omega^{-4/3}$. For the linearly polarized
case ($e=1$), experiments~\cite{galv88} and numerical
computations~\cite{jens89} show that for $F_0=F_c n^4> 0.02$ (when
$n$ is the principal quantum number) there is ionization for
$\Omega_0=1/\omega$ (the scaled frequency) close to 1. Here we find
agreement with this value: for $F_0>0.022$, there is no invariant
curve in phase space and diffusion can occur. For the circularly
polarized case ($e\not= 1$), the conclusions are less clear. However
since there is a broad stable region $e\in [0,0.8]$, we expect the
circularly polarized driven atoms to ionize less easily than the
linearly driven ones. Since the orbits of high eccentricities are the
ones which ionize, we expect the behavior of ionization curves for
circularly polarized microwaves to be similar to the ones for linearly
polarized ones. This is consistent with experiments (Fig.~1 of
Ref.~\cite{bell96}).

\paragraph*{Progress to chaos through noble tori--}

The renormalization map allows us to determine accurately the
frequency of the last invariant torus to break-up.  The conventional wisdom is that
the last invariant torus surviving with increasing the amplitude of
the perturbation is the one with frequency equal to the golden mean
$\gamma=(\sqrt{5}-1)/2$. The example we
study shows that this belief is mistaken. Figure~4 shows the value of the
frequency of the last invariant torus between resonances 1:1 and 2:1, and
between 2:1 and 3:1 (for co-rotating and counter-rotating orbits), as a function
of the parameter $e$. We have identified the frequency of these tori (by
accurate computation of critical couplings in the neighborhood of
these frequencies). In the range of the parameter $e$ and for these
regions of phase space, each last invariant torus is a noble one in
the sense that its frequency $\omega$ is equivalent to the golden mean
$\gamma$, i.e.\ there exist integers
$(a,b,c,d)\in {\mathbb{Z}}^4$ such that $ad-bc=\pm 1$ and
$\omega=(a\gamma+b)/(c\gamma+d)$ (or equivalently the tail of the
continued fraction expansion of $\omega$ is a sequence of 1). For
example, for $e=0.45$, the last invariant torus is expected to be
$(\gamma+1)/(2\gamma+2)\approx 0.7236$ (see Fig.~1). For the region of high
eccentricities (which ionize more easily), the expected frequency for
the last invariant torus is $\omega=(2\gamma+1)/(3\gamma+2)\approx
0.7835$. The observation that the last invariant torus is noble has
also been checked for the counter-rotating case. Whether or not this
observation holds for a generic Hamiltonian remains an open
question~\cite{mack88}.

\paragraph*{Conclusion--}
Using the renormalization method, we find that the
empirical 2/3-rule Chirikov criterion is surprizingly accurate for the
onset of chaos in the stochastic ionization problem. The model studied in this
Letter, the hydrogen atom driven by strong microwaves, shows how renormalization
and empirical rules can be used together in order to obtain very accurate
information on the stability of the system.


\newpage

\begin{figure}          
\centerline{ \includegraphics[scale=0.5]{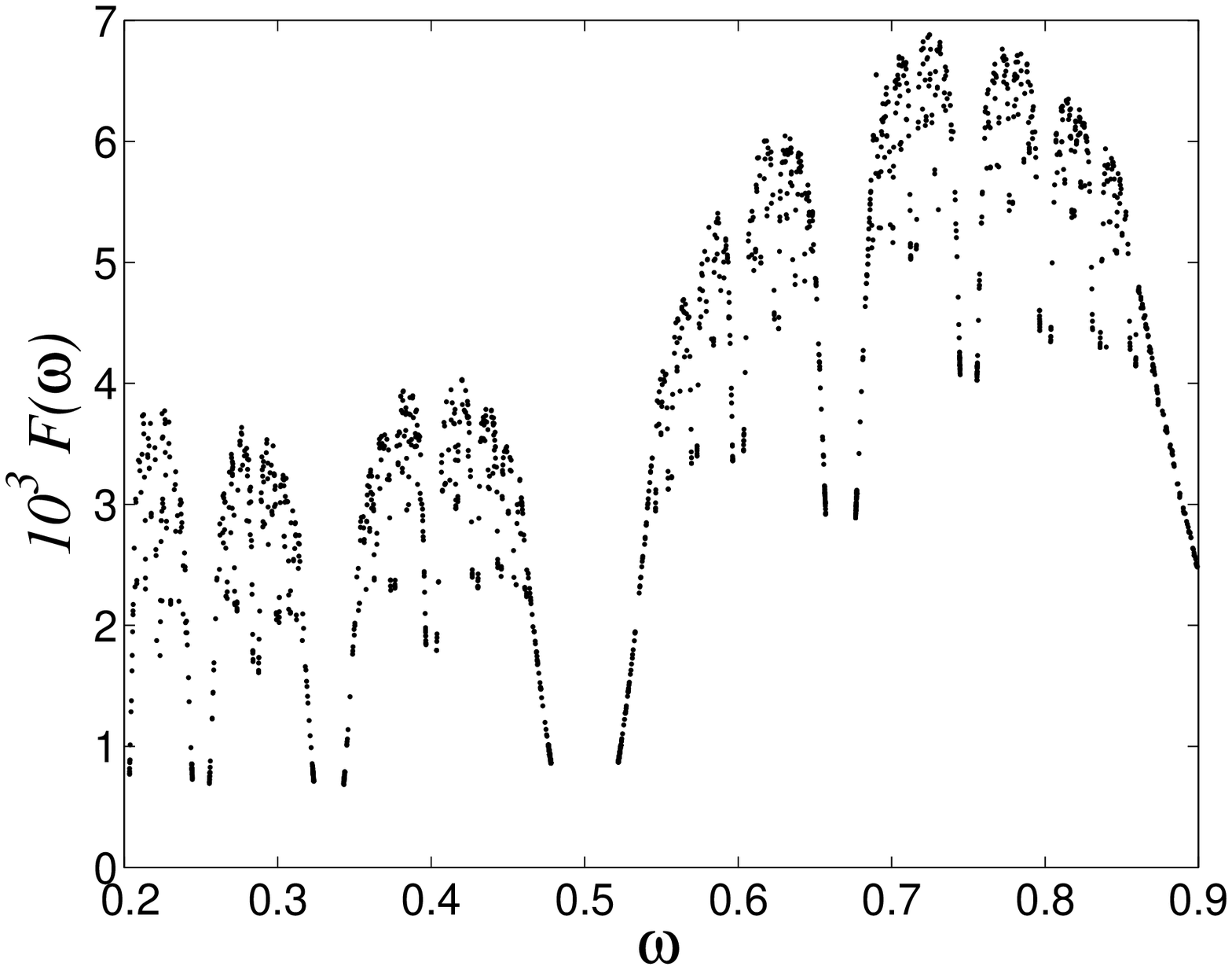}}
\caption{Critical function $F_c(\omega)$ for $\omega\in [0,1]$ for
Hamiltonian~(\ref{eqn:hamresc}) with $e=0.45$.} 
\label{fig:1}
\end{figure}

\newpage

\begin{figure}
\centerline{\includegraphics*[scale=0.5]{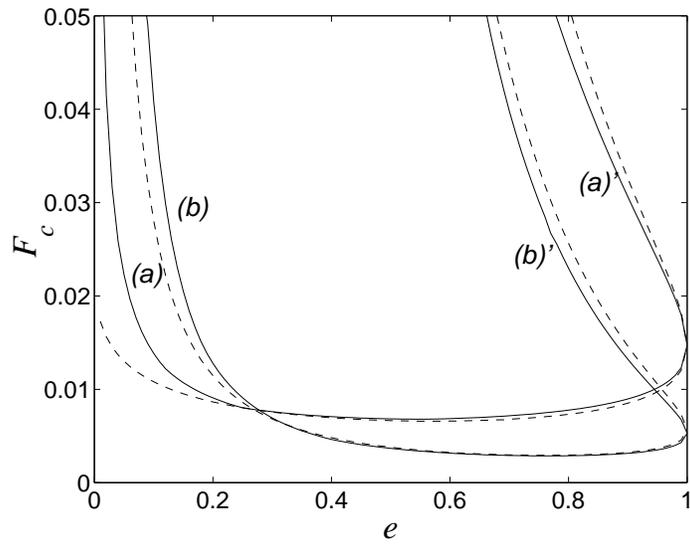}}
 \caption{Critical threshold $F_c$ between $(a)$ resonances 1:1 and
2:1 and $(b)$ resonances 2:1 and 3:1 for co-rotating $(a)$, $(b)$ and
counter-rotating $(a)'$, $(b)'$ orbits. The continuous curves are obtained by
renormalization and dashed curves by the 2/3-rule criterion.}
\label{fig:2}
\end{figure}

\newpage

\begin{figure}
\centerline{\includegraphics*[scale=0.4,angle=-90]{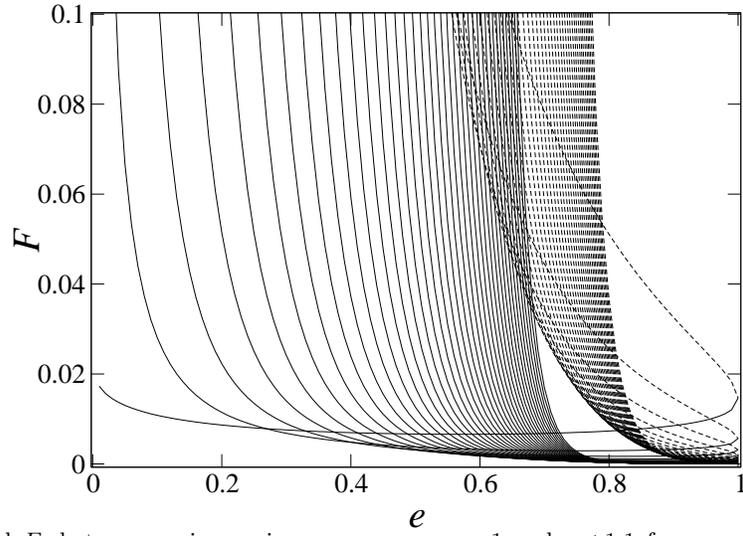}}
\caption{Critical threshold $F_c$ between various primary
resonances $m$:1 and $m$+1:1 for $m=1,\ldots,40$ from left to right,
obtained by the 2/3-rule criterion for co-rotating orbits
(continuous curves) and counter-rotating orbits (dashed curves).}
\label{fig:3} 
\end{figure}

\newpage

\begin{figure}
\centerline{
\includegraphics*[scale=0.5]{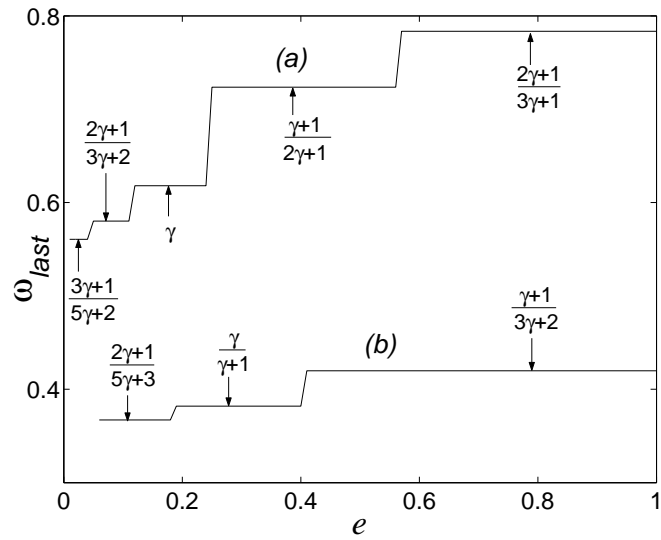}}
 \caption{Value $\omega_{last}$ of the frequency of the last invariant
torus for Hamiltonian~(\ref{eqn:hamresc}) as a function of the
parameter $e$~: $(a)$ between resonances 1:1
and 2:1, and $(b)$ between resonances 2:1 and 3:1.} \label{fig:4}
\end{figure}

\end{document}